\documentclass{article}
\usepackage{times}
\usepackage{geometry}
\usepackage{titlesec}
\usepackage{setspace} % For adjusting line spacing
\usepackage{ragged2e}
\usepackage{hyperref}
\usepackage{graphicx}
\usepackage{caption}
\usepackage{hhline}
\usepackage{fancyhdr}
\usepackage{ifthen}
\usepackage{url}
\usepackage{footmisc}
\usepackage{amsmath,amssymb}
\usepackage{xcolor}
\usepackage{subfigure}
\usepackage{etoolbox}
\usepackage[style=numeric, sorting=none]{biblatex}
\usepackage{textcase}
\usepackage{tabularx} % For better control over table column widths
\usepackage{comment}

\definecolor{lightgraytext}{gray}{0.4} % Adjust the shade of grey as needed

\DeclareCaptionLabelFormat{figlabel}{{FIG. #2. }}
\DeclareCaptionLabelFormat{tablelabel}{{TABLE #2. }}
\titleformat{\section}
  {\fontsize{10}{12}\selectfont\MakeUppercase} % Set font size and style
  {\thesection.} % Section number
  {0.5em} % Horizontal spacing
  {} % Code before title

\titleformat{\subsection}
  {\fontsize{10}{12}\selectfont\bfseries} % Set font size and style
  {\thesubsection.} % Section number
  {0.5em} % Horizontal spacing
  {} % Code before title

\titleformat{\subsubsection}
  {\fontsize{10}{12}\selectfont\itshape} % Set font size and style
  {\thesubsubsection.} % Section number
  {0.5em} % Horizontal spacing
  {} % Code before title

\newcommand{\centeredparheader}[1]{%
    \textcolor{lightgraytext}{%
        \parbox[c]{\textwidth}{%
            \centering #1%
        }%
    }%
}

\fancypagestyle{plain}{% This is for the first page
  \fancyhf{}%
  \fancyhead[C]{\centeredparheader{%
    \fontsize{8}{0}{Carey et al.} }%
}}

\newcommand{\evenheader}{\centeredparheader{%
   \fontsize{8}{0}{IAEA-CN-2605}}}
    
\newcommand{\oddheader}{\centeredparheader{%
    \fontsize{8}{0}{Carey et al.}}}
    
\begin{document}

\title{\raggedright
  \textbf{\fontsize{12}{15}\selectfont
  \MakeUppercase{DATA EFFICIENCY AND LONG-TERM PREDICTION CAPABILITIES FOR NEURAL OPERATOR SURROGATE MODELS OF CORE AND EDGE PLASMA CODES
}}
  \vspace{-3.2em}
  \setstretch{0.6}
}
\date{}
\maketitle
%\subtitle{\raggedright\textbf{\textit{{\fontsize{14}\ Subtitle if needed in Times New Roman 12 point bold italic, sentence case}}}\vspace{2em}}

\begin{flushleft}
  N. Carey \\
  UKAEA, Culham Centre for Fusion Energy \\
  Abingdon, UK \\
  Email: naomi.carey@ukaea.uk
\end{flushleft}

\begin{flushleft}
  L. Zanisi \\
  UKAEA, Culham Centre for Fusion Energy \\
  Abingdon, UK \\
  Email: lorenzo.zanisi@ukaea.uk
\end{flushleft}

\begin{flushleft}
 S. Pamela \\
  UKAEA, Culham Centre for Fusion Energy \\
  Abingdon, UK \\
\end{flushleft}

\begin{flushleft}
 V. Gopakumar \\
  UKAEA, Culham Centre for Fusion Energy \\
  Abingdon, UK \\
\end{flushleft}

\begin{flushleft}
  J. Omotani \\
  UKAEA, Culham Centre for Fusion Energy \\
  Abingdon, UK \\
\end{flushleft}

\begin{flushleft}
  J. Buchanan \\
  UKAEA, Culham Centre for Fusion Energy \\
  Abingdon, UK \\
\end{flushleft}

\begin{flushleft}
  J. Brandstetter \\
  Microsoft Research AI4Science \\
  Amsterdam, The Netherlands \\
\end{flushleft}

\begin{flushleft}
  the JOREK team 
\end{flushleft}

\section*{\bfseries\MakeTextUppercase{a}\MakeTextLowercase{bstract}}
{\fontsize{9pt}{12pt}\selectfont\hspace{1cm}Simulation-based plasma scenario development, optimization and control are crucial elements towards the successful deployment of next-generation experimental tokamaks and Fusion power plants. Current simulation codes require extremely intensive use of HPC resources that make them unsuitable for iterative or real time applications. Neural network based surrogate models of expensive simulators have been proposed to speed up such costly workflows. Current efforts in this direction in the Fusion community are mostly limited to point estimates of quantities of interest or simple 1D PDE models, with a few notable exceptions. While the AI literature on methods for neural PDE surrogate models is rich, performance benchmarks for Fusion-relevant 2D fields has so far remained flimited. In this work neural PDE surrogates are trained for the JOREK MHD code and the STORM scrape-off layer code using the PDEArena library\footnote{\url{https://github.com/microsoft/pdearena}}. The performance of these surrogate models is investigated as a function of training set size as well as for long-term predictions. The performance of surrogate models that are trained on either one variable or multiple variables at once is also considered. It is found that surrogates that are trained on more data perform best for both long- and short-term predictions. Additionally, surrogate models trained on multiple variables achieve higher accuracy and more stable performance. Downsampling the training set in time may provide stability in the long term at the expense of the short term predictive capability, but visual inspection of the resulting fields suggests that multiple metrics should be used to evaluate performance.
    }

\renewcommand{\footnoterule}{\hrule width \linewidth}

\section{Introduction}
A crucial step for fusion research to build commercial fusion devices is being able to model plasma behaviour. Modelling of both the scrape-off layer (SOL) and the core plasma is fundamental to ensure appropriate divertor and core performance, and their reliable characterisation is desirable for not just designing and interpreting experiments in the current generation of reactors, but to inform the next generation too (i.e., ITER and STEP \cite{Loarte2007Chapter4P,Eich2013ScalingOT}).
Plasma behaviour can be modelled as a strongly coupled system of partial differential equations (PDEs). Obtaining numerical solutions of these PDEs is computationally very intensive due to the spatial and temporal scales involved. Existing frameworks used for modelling, such as BOUT++ \cite{Dudson_2009}, implement numerical solvers with a trade-off between accuracy and speed. Even for reduced-order models, which are often 1 dimensional (e.g., DIV1D \cite{Derks2022}), the computational cost is still high or even prohibitive for iterative applications, such as optimisation, or real-time control. 
Neural network (NN) based surrogate models offer a promising avenue to enable the fast estimation of plasma states given the initial and boundary conditions for a physical model of choice. For example, Convolutional Neural Networks (CNNs) have been applied to emulate the behaviour of SOLPS \cite{WIESEN2015480,Gopakumar2020,DASBACH2023101396} over a restricted area of its parameter space, demonstrating orders of magnitude speedup. Nevertheless, solutions derived from CNNs lack discretization invariance, and therefore are applicable only to the specific discretization of the numerical PDE solution that they were trained on. Recent work has introduced Neural Operators (NOs, e.g., \cite{kovachki2023neural}), a family of surrogate models that learn infinite-dimensional mappings between function spaces, and that are thus capable of learning a continuous, discretization-invariant representation of PDE solutions. The Fourier Neural Operator (FNO), where convolutional filters are learned in Fourier space, has proved a particularly successful architecture for modelling PDEs with NNs \cite{li2021fourier}.
 Fusion applications of NOs have so far been sparse. A first exploration of NOs for emulation of MHD was presented in \cite{gopakumar2023plasma}, where it was shown that the FNO outperforms non discretization-invariant, CNN-based architectures. The FNO was also capable of accurately predicting the evolution of plasma in the Mega Ampere Spherical Tokamak visible camera images. \cite{Poels_2023} have benchmarked the performance of a range of NOs at reproducing DIV1D simulations, and demonstrated that the FNO provides a competitive baseline. 
This work will present the results of work to develop fast FNO-based surrogate models of simulations from STORM \cite{Walkden2016-ys} as well as JOREK \cite{Hoelzl2021-ug}. Built on the BOUT++ framework, STORM is a  fluid code focused on modelling turbulence and transport processes in the scrape-off layer region (SOL) and has been used to carry out nonlinear flux-driven simulations in double null tokamak geometry with realistic parameters \cite{Riva2019-fh}. JOREK is a continuously developed simulation code that is widely used to study large-scale plasma instabilities from the core and edge regions \cite{Smith2020-cv}. This work will make use of the PDEArena library \cite{gupta2022multispatiotemporalscale}, which is built with benchmarking different neural PDE surrogate architectures an easy task. As a first proof of concept, this paper will focus only on the FNO and further methodologies will be explored in future work.

This work focuses in particular on the following issues:
\begin{itemize}
    \item The size of the training dataset required to obtain performing surrogate models. As NOs are trained on numerical solutions of PDEs, which can be expensive to obtain, this is an important efficiency metric. 
    \item The performance of the surrogate models for long term prediction. This is prone to degradation due to errors accumulating and eventually driving the NO outside of the training distribution. A strategy is experimented with where the NO is trained on progressively coarser time resolutions.
    \item The performance of the surrogate models in cases where multiple variables are learned concurrently by the same model, thus exploiting the correlations induced by the physical models, versus the naive case where surrogates are learned on individual variables.
\end{itemize}

This paper is structured as follows: Section 2 provides a description of the datasets, Section 3 introduces the NO training configuration, Section 4 outlines the results of the experiments. The key results are summarised in section 5, which includes also a discussion on the future extensions of this work.

\section{Data}
\label{sec:data}
Training datasets are created from STORM and JOREK simulations. JOREK , while STORM evolves density and electric potential for 1000 timesteps. Both simulations are run in 2D rectangular slab geometry for the purposes of these experiments, however the long term goal will be to expand to the 3D case with more complex simulations in full toroidal geometry.
The JOREK dataset is generated using a simplified MHD simulation model evolving density, temperature and electric potential, and involving the radial convection of multiple plasma blobs. The simulations were all run in 2D square 100x100 slab geometry with the width and height being 1m, centred at major radius R=10m. The plasma blobs were initialised on top of a low background density and temperature, and the pressure gradients of the blobs cause a buoyancy-like effect that leads them to move outwards due to the toroidal geometry. The “multiple blobs with non-uniform temperature” dataset is used from \cite{gopakumar2023plasma}, which comprises 2000 simulations with different blob numbers, starting locations, temperature fluctuations and density. Further details can be found in \cite{gopakumar2023plasma}. 1800 simulations are used for training and 200 for testing, unless stated otherwise. The JOREK simulations were run for 2000 timesteps, with a very small timestep size to ensure stability. In \cite{gopakumar2023plasma} the time domain was then downsampled by a factor of 10, for a total of 200 timesteps available, and the same approach will be adopted here.

The STORM simulations evolve density and electric potential. A vertical band of density source generates fluid turbulence in the radial direction due to the toroidal geometry. The dataset is obtained by varying the amplitude and width of the source. Throughout the time steps, density at the source is increased as a function of time in order to maintain density at the starting band throughout the trajectory. The simulations were all run in 2D regular slab geometry (384x256, scaled down by a half to reduce training time) with the dimensions being 150 and 100mm, located around the separatrix, allowing for analysis of much smaller scale turbulence. 1000 STORM simulations were generated, of which 900 were used for training and  100 for testing, unless stated otherwise. For STORM, a total of 1000 timesteps were run and only post saturation of turbulence was considered, which was identified as when vorticity (which is derived from the potential) started increasing from an initial value of zero. This was done as early experiments showed that the transition between the two regimes was extremely difficult to capture accurately.

Since both datasets involved the gradual diffusion of inhomogeneous density, the data distribution within the spatial domain was severely imbalanced. Normalisation of the dataset was performed in order to address this, specifically the training dataset was scaled linearly to allow for the field values to lie between -1 and 1 and the same transformation was performed on the evaluation dataset.

\section{Methods}
\label{sec:surrogates}
The PDEarena platform \cite{gupta2022multispatiotemporalscale} includes several options of Neural PDE solvers, including the Fourier-Neural-Operator method (FNO) \cite{li2021fourier} used in this work. The FNO architecture consists of stacked blocks defined as follows,
\begin{equation}
    y = \sigma\bigg(\mathcal{F}^{-1}\big(R\mathcal{F}(x)\big) + Wx + b\bigg)
\end{equation}
which, for an input $x$ combine a learned Fourier representation $R$ and a linear transformation of the field $W$ with bias $b$ using a nonlinearity $\sigma$. Here, $\mathcal{F}$ is the Fourier transform. See \cite{li2021fourier} for more details.
 
PDEarena was used as a code base and extended to support STORM and JOREK simulation data. For the model, a modified version of the FNO configuration “FNO-128-32m” in PDEarena is used with the number of Fourier blocks increased to 3, and where the grid discretisation is concatenated in the same dimension as field data where available as it provides improvement in performance as demonstrated in \cite{gopakumar2023plasma}.

\begin{figure}
    \centering
    \includegraphics[width=0.6\textwidth]{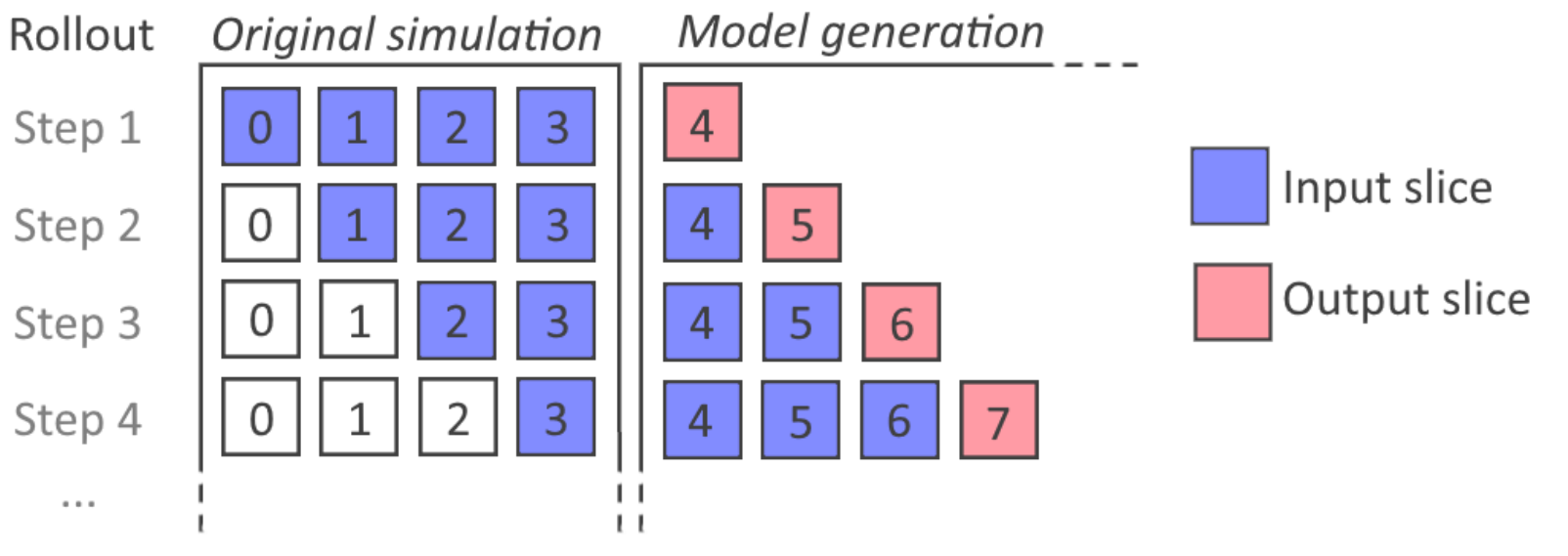}
    \caption{Example rollout using surrogate where the surrogate takes in 4 and generates 1 timestep. For the initial rollout step, the surrogate uses only timesteps from the original simulation, but rollout steps beyond that start using an increasing portion of generated rollout steps.
}
    \label{fig:rollout}
\end{figure}

The FNO takes in field values across a 2D grid for an initial set of time steps and outputs a set of later time steps across the same grid. An input feed of 20 time steps of field information and an output of 5 time steps was arbitrarily chosen. To achieve longer rollouts, the output (coupled with later timesteps from the input where necessary) is fed back in the FNO to obtain further field evolution in time and each iteration is called a rollout step (see Figure \ref{fig:rollout}).
Each FNO was trained on a single rollout step with a random starting point. Likewise, validation was performed using a single rollout step, and an early stopping strategy with patience of 10. All results shown are for models trained on solely the density variable apart from the comparison between individual vs multivariable models (see Section \ref{sec:individual_mv}). Each model was trained for up to 72 hours using the Adam optimizer with cosine annealing learning rate scheduler with the starting learning rate of 0.0002 and minimum learning rate value of 1.e-7 for both. Hyperparameter optimisation is expensive and has not been performed in this case,  but will be considered in further studies. The model was trained on a mean squared error (MSE) loss function and evaluated on mean absolute error (MAE). The model output is evaluated for the normalised dataset, in order to allow for fair comparisons between simulation models and corresponding variables.

\section{Outcomes}
\label{sec:Results}
\subsection{Initial results}
\label{sec:initial_results}
\begin{figure}
    \centering
    \subfigure[]{
         \includegraphics[width=0.85\textwidth]{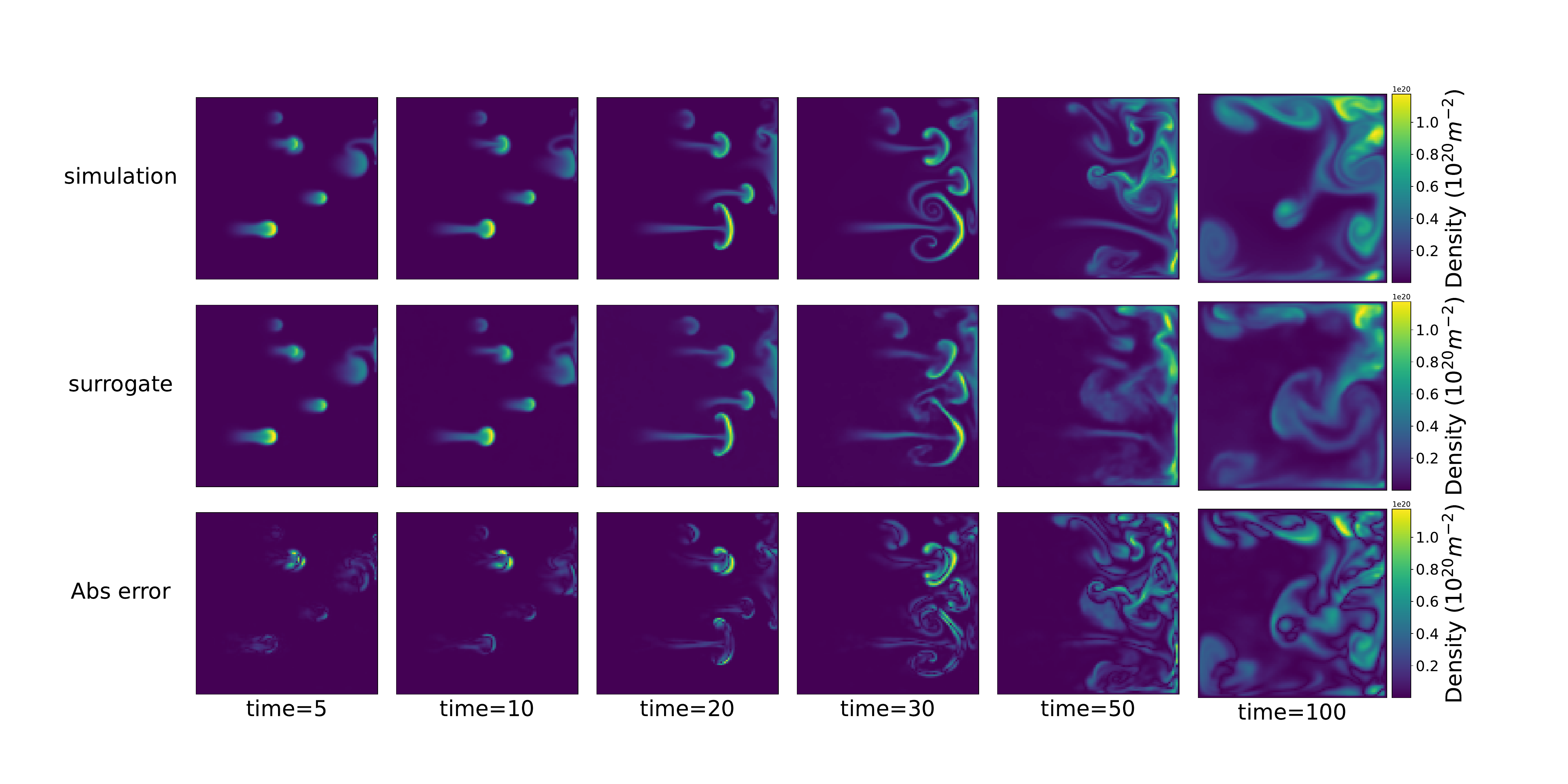}
    }%
     \hspace{0.1\linewidth} 
    \subfigure[]{
         \includegraphics[width=0.95\textwidth]{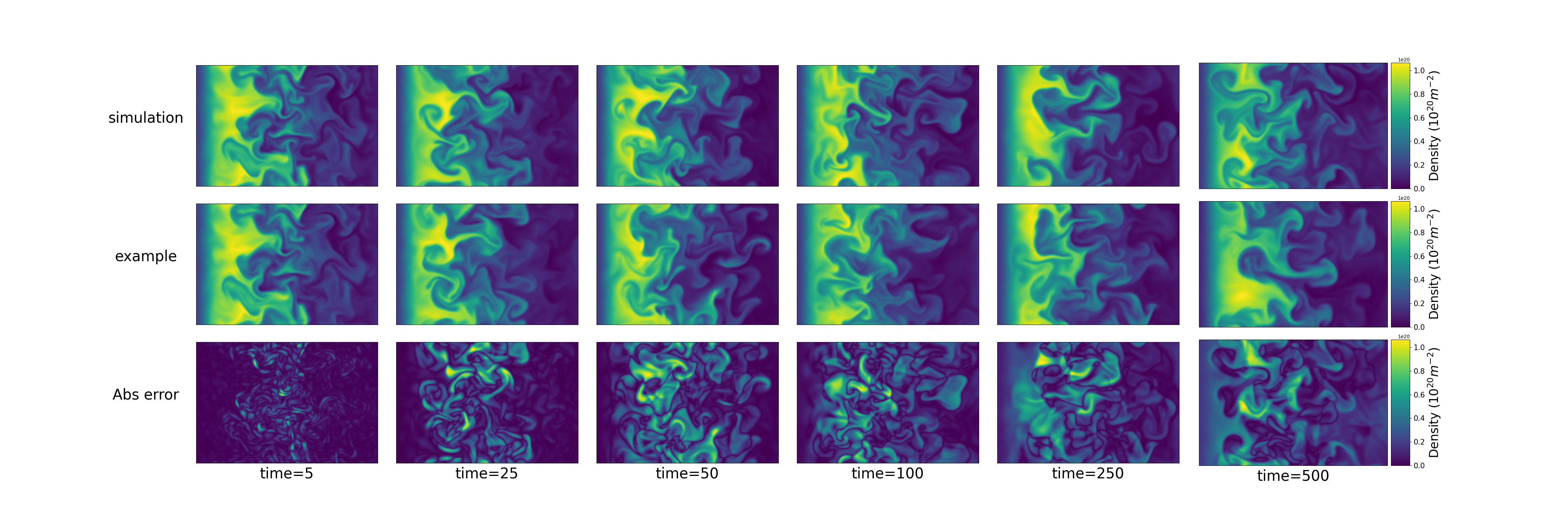}
    }    
    \caption{FNO model rollout plotted at specific time steps, for an exmaple JOREK run (a), and an example STORM run (b), for the electron density.
}
    \label{fig:initial_results}
\end{figure}

As a first experiment, two FNOs were trained to learn the evolution of the density field from 1900 JOREK simulations and 900 STORM simulations respectively, converging to a test MSE of 3.14e-4 and 2.02e-4 respectively. Rollouts for test samples are shown in Figure \ref{fig:initial_results} and it can be seen that both surrogates were able to learn global features, such as the location of the blobs in JOREK (Figure \ref{fig:initial_results} upper panels) and the location of the SOL (Figure \ref{fig:rollout}, bottom panels).
However, it is clear that for longer time rollouts the finer details of the fields are quite different from the ground truth. This is shown quantitatively in Figure \ref{fig:dataset_size} for the density fields, and in the remainder of the paper for the other variables. A combination of two  factors might result in the observed behaviour:
\begin{itemize}
    \item Model error: the nature of data driven approaches implies that  the surrogate model learns an approximation of future plasma states. The autoregressive strategy results in such approximations being used to predict future timesteps, which results in compounding of error and in the model being deployed outside of the distribution of data it was trained on. 
    \item Physics error: the chaotic nature of the turbulence  implies that even if the surrogate model almost perfectly replicates turbulence physics, small, often unavoidable errors at the beginning of the rollout will result in a large error towards the end. 
\end{itemize}

\subsection{Dataset size and long rollouts}
\label{sec:dataset_size}
\begin{figure}
    \centering
    \subfigure[]{
        \includegraphics[width=0.4\textwidth]{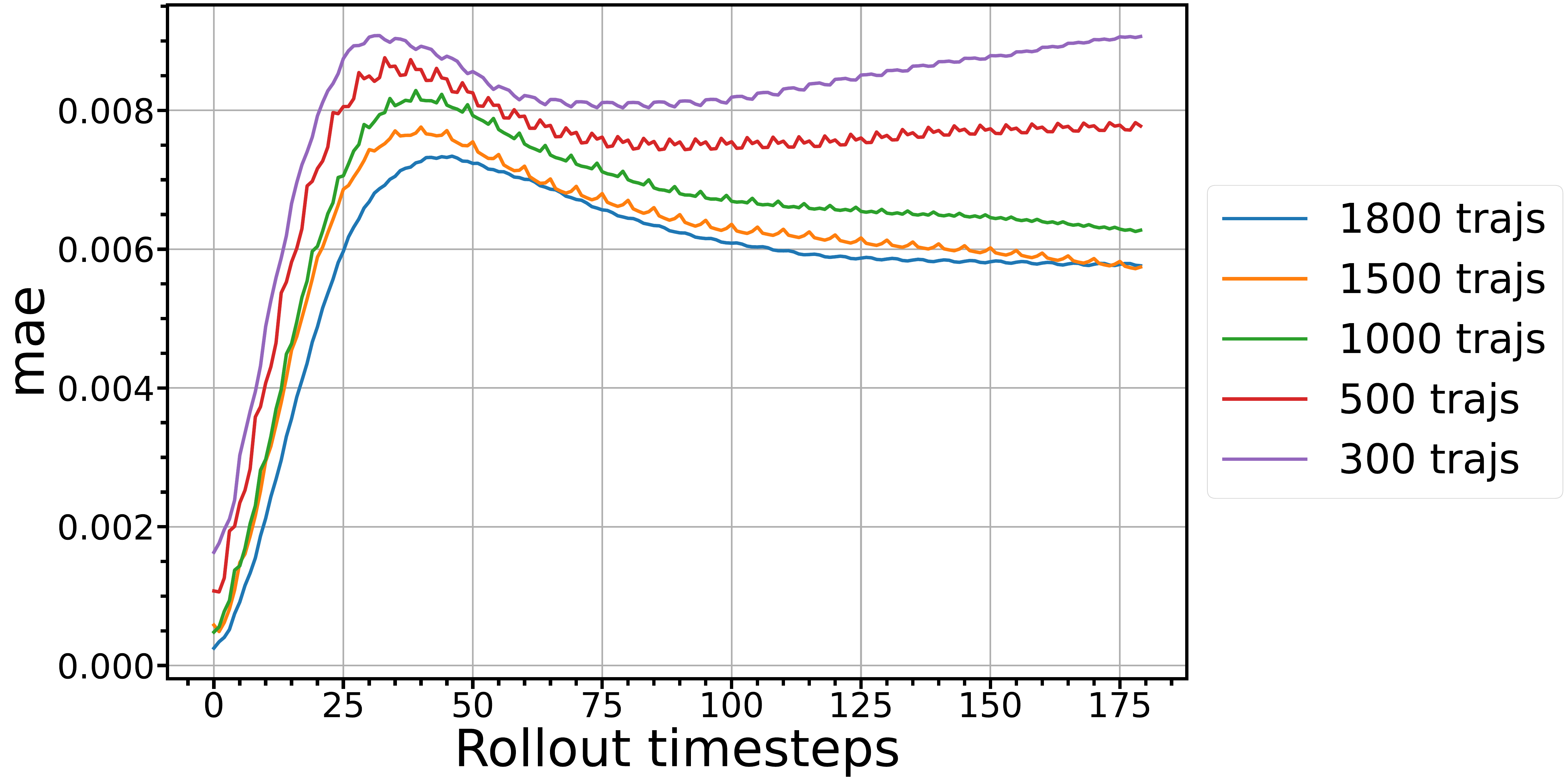}
    }
    \subfigure[]{
        \includegraphics[width=0.4\textwidth]{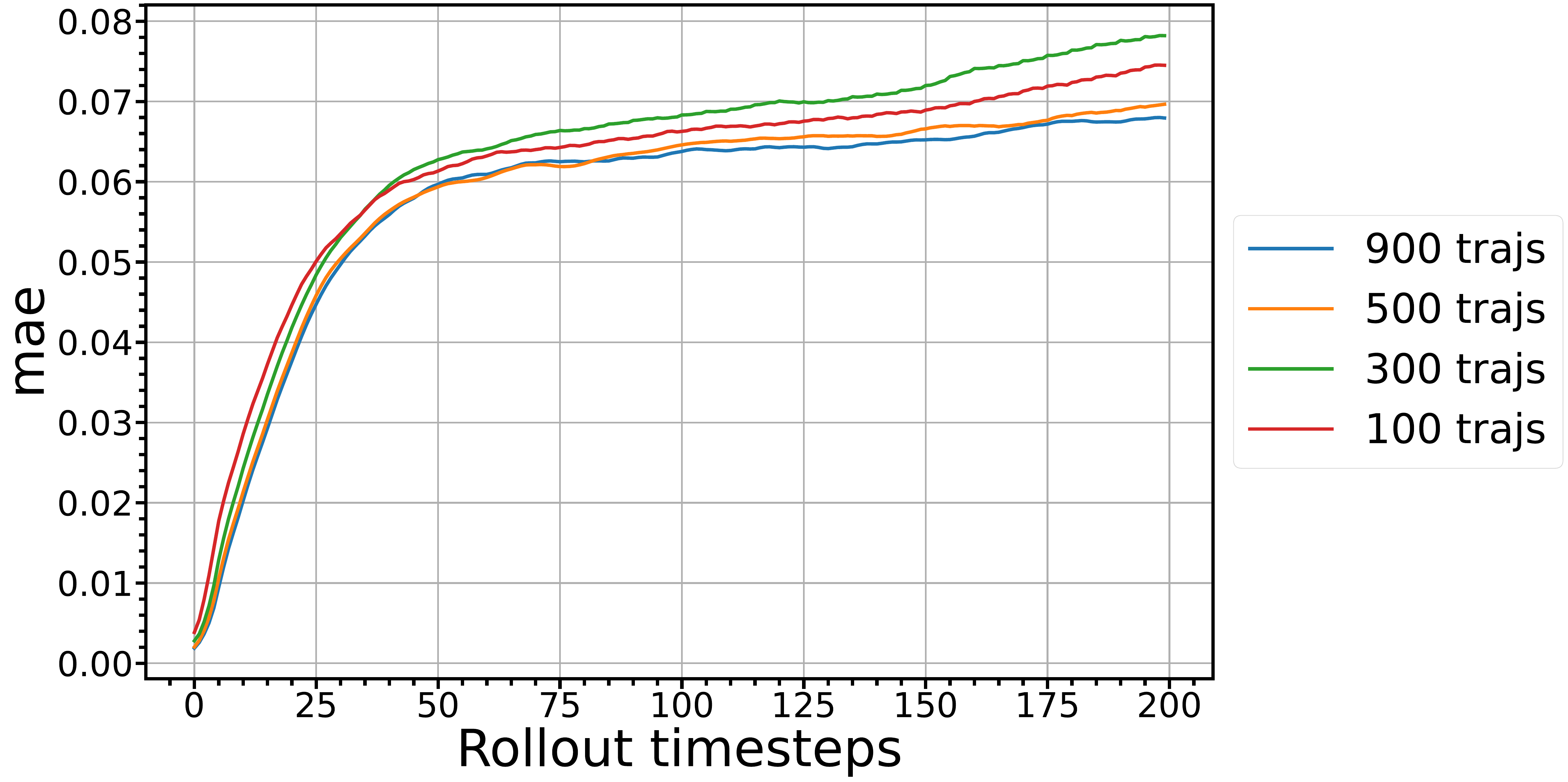}
    }    
    \caption{Impact of training dataset size on model rollout performance for JOREK (a) and STORM (b)}
    \label{fig:dataset_size}
\end{figure}

The next experiment investigates the impact of the training dataset size on the performance of the FNO. For this purpose, different sized subsets of the original training dataset was used, leaving the validation dataset the same for all models. Multiple FNO models with training dataset size that varies from 900 to 300 STORM simulations and 1800 to 300 JOREK simulations are trained. All the models are constructed and trained with identical hyperparameters with the exception of the training dataset size.
As seen in Figure \ref{fig:dataset_size}), in general, the larger the training dataset, the better the performance of the model.

\subsection{Individual vs multivariable}
\label{sec:individual_mv}
\begin{figure}[!h]
    \centering
    \subfigure[]{
        \includegraphics[width=0.45\textwidth]{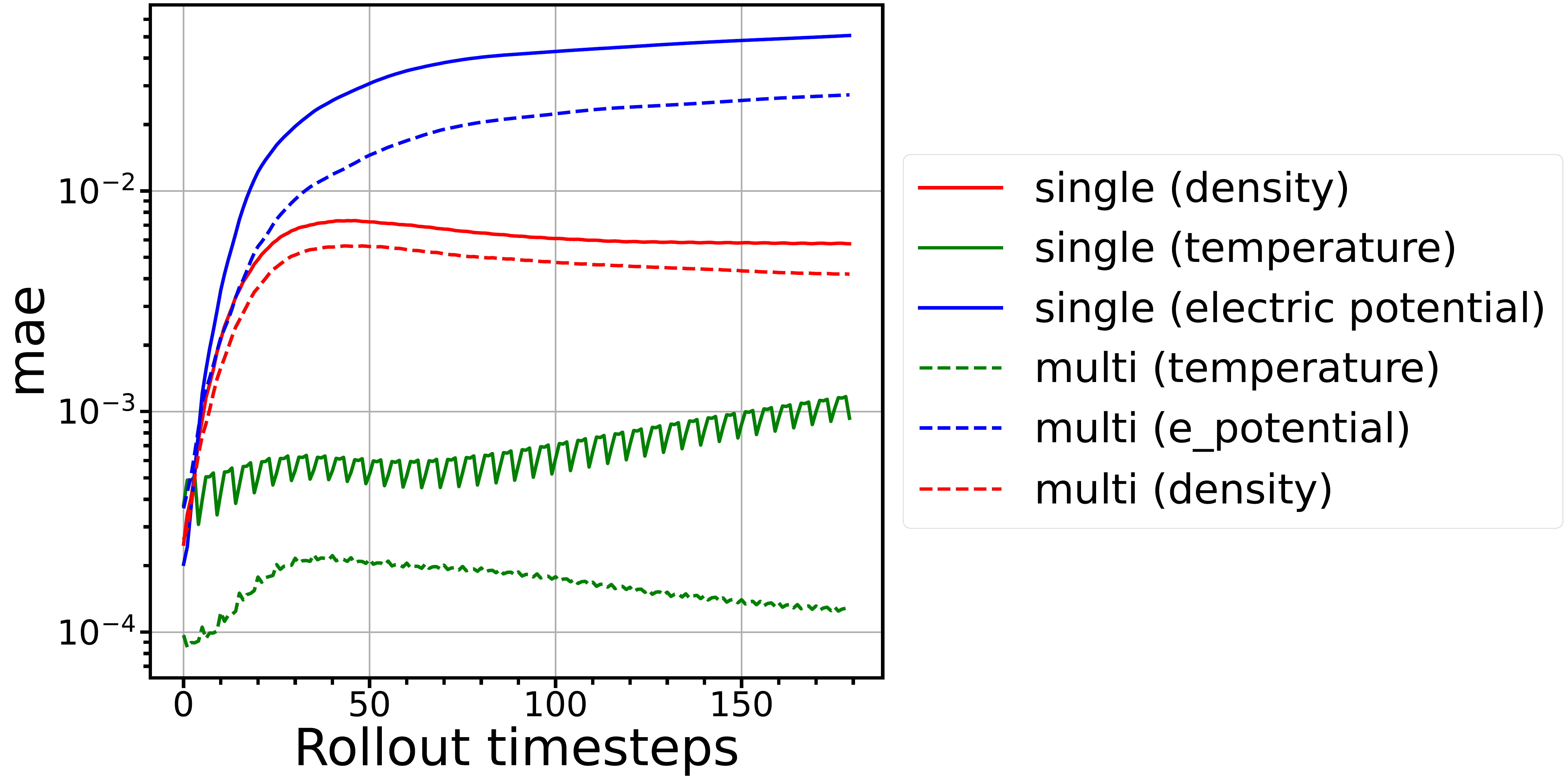}
    }    
    \subfigure[]{
        \includegraphics[width=0.45\textwidth]{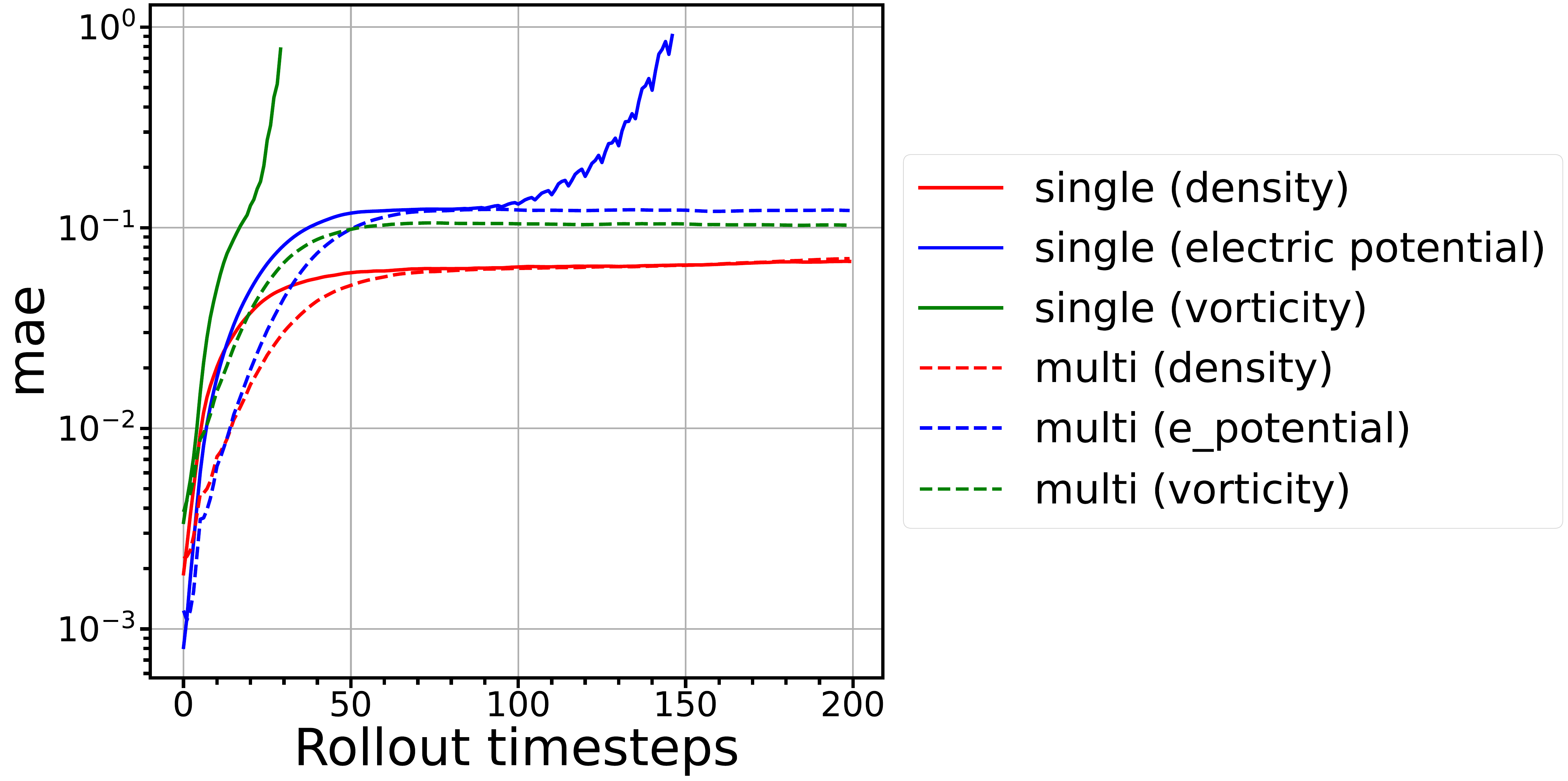}    }        
    \caption{Comparing the error growth of different variables across time roll-out of the Individual FNOs (Single) and the Multi-variable FNO (Multi) for JOREK (a) and STORM (b).}
    \label{fig:individual_mv}
\end{figure}

For both sets of simulations, evolution of the plasma state requires modelling tightly coupled variables as prescribed by the physics of the systems of PDEs of interest. The multi-variable FNO in PDEarena was used as described in Section \ref{sec:surrogates}  to test whether modelling multiple variables together is beneficial in the neural operator framework. As shown in Figure \ref{fig:individual_mv}, the multi-variable FNO achieves better performance in all variables for both the JOREK and STORM simulations and this holds true for the entire rollout, while for some STORM surrogates trained on individual variables the predictions even diverge. The conclusion drawn from this is that for highly-correlated fields  in simulations such as STORM and JOREK, the multi-variable FNO is able to better capture the dynamics compared to the case of individual variables modelled separately, which allows an improved handling of  the accumulated rollout error for a fixed choice of hyperparameters. 

\subsection{On time subsampling for long term predictions}
\label{sec:time_subsampling}
The rollout strategy of evaluation is very desirable as it allows for flexibility in the number of output timesteps. However, as seen in the previous sections, this strategy is prone to performance degradation throughout the rollout. In order to improve long term rollout performance, an experiment where models are trained on sparser timesteps is conducted. For this purpose, the data is downsampled in time by a factor of N. By training on sparser timesteps, the model is trained on the evolution of variables for longer periods of time which, it is hoped, will result in a slower rate at which errors are integrated into the input, and therefore propagated forward in time via the autoregressive strategy. Specifically, every N=1, 2 and 10 timesteps is used for the model trained on STORM simulations and 1, 2 and 5 for the model trained on JOREK due to limitations associated with trajectory length available. With the exception of the change above, all models are constructed and trained in identical conditions.

\begin{figure}[!h]
    \centering
    \subfigure[]{
        \includegraphics[width=0.45\textwidth]{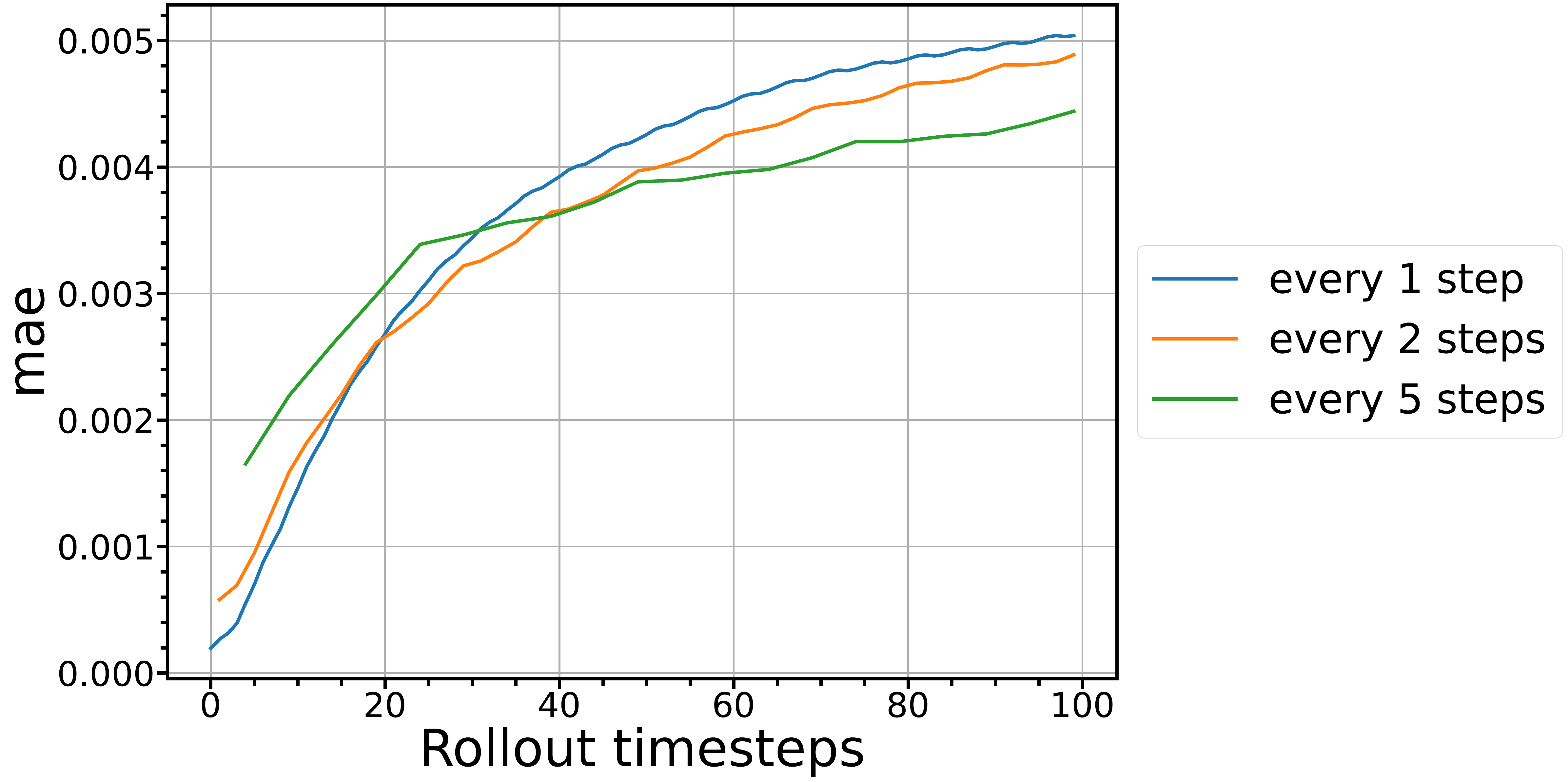}
    }    
    \subfigure[]{
        \includegraphics[width=0.45\textwidth]{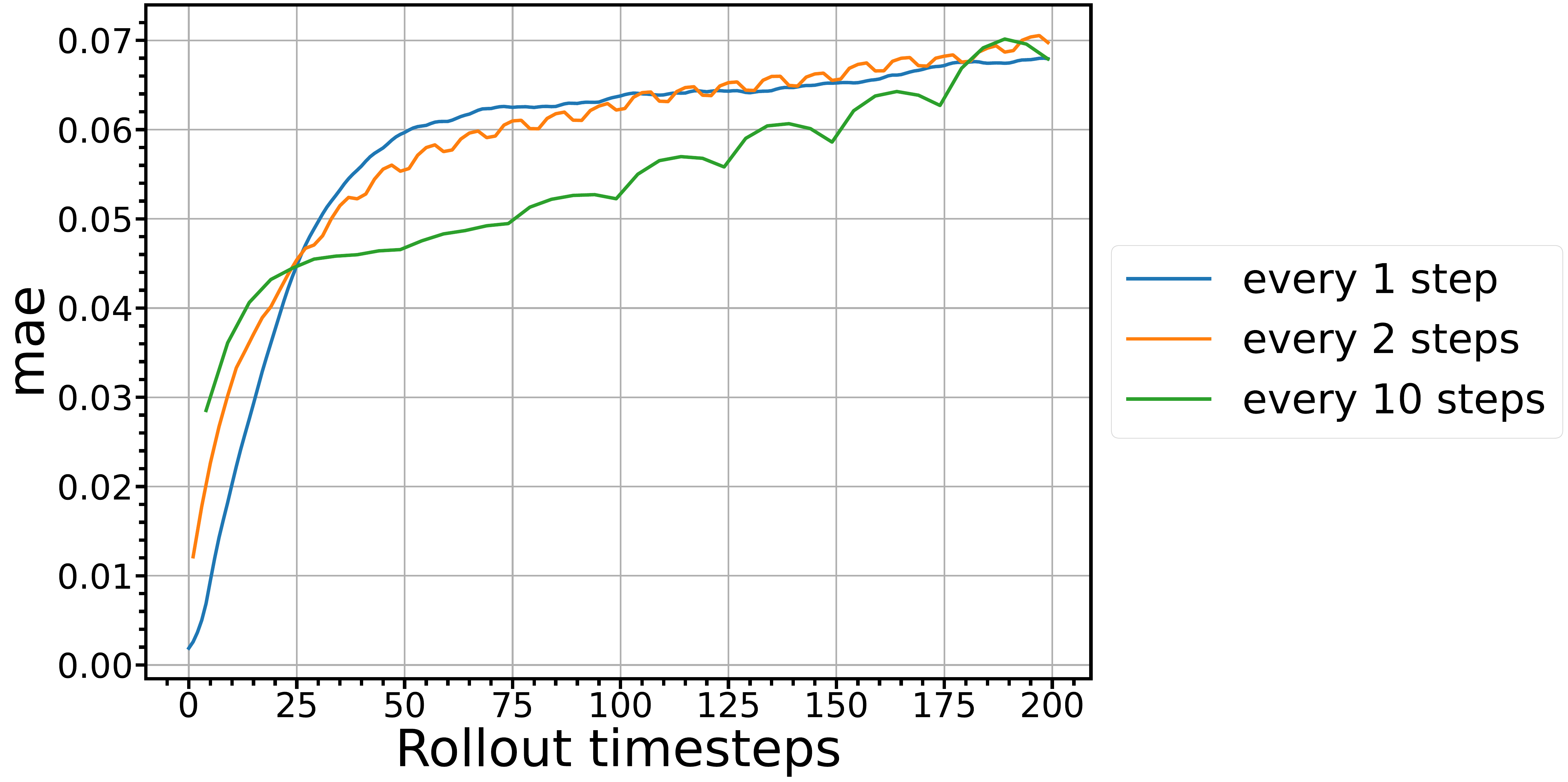}
    }        
    \caption{ Impact of time downsampling during training on model rollout performance for up to 100 timesteps in JOREK (a) and 200 in STORM (b).
}
    \label{fig:time_subsampling_1}
\end{figure}

\begin{figure}[!h]
    \centering
    \subfigure[]{
        \includegraphics[width=0.85\textwidth]{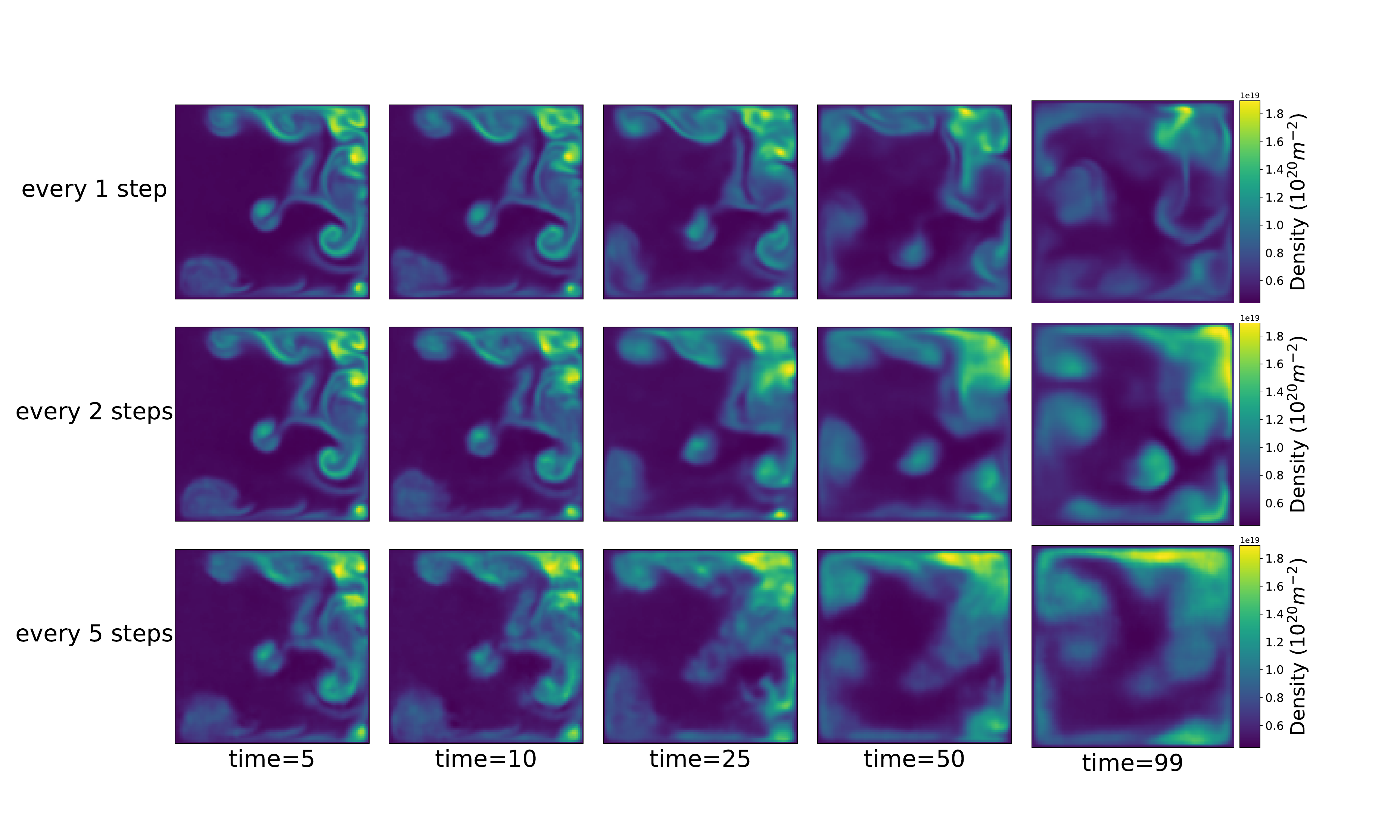}
    }        
     \hspace{0.1\linewidth}     
    \subfigure[]{
        \includegraphics[width=0.95\textwidth]{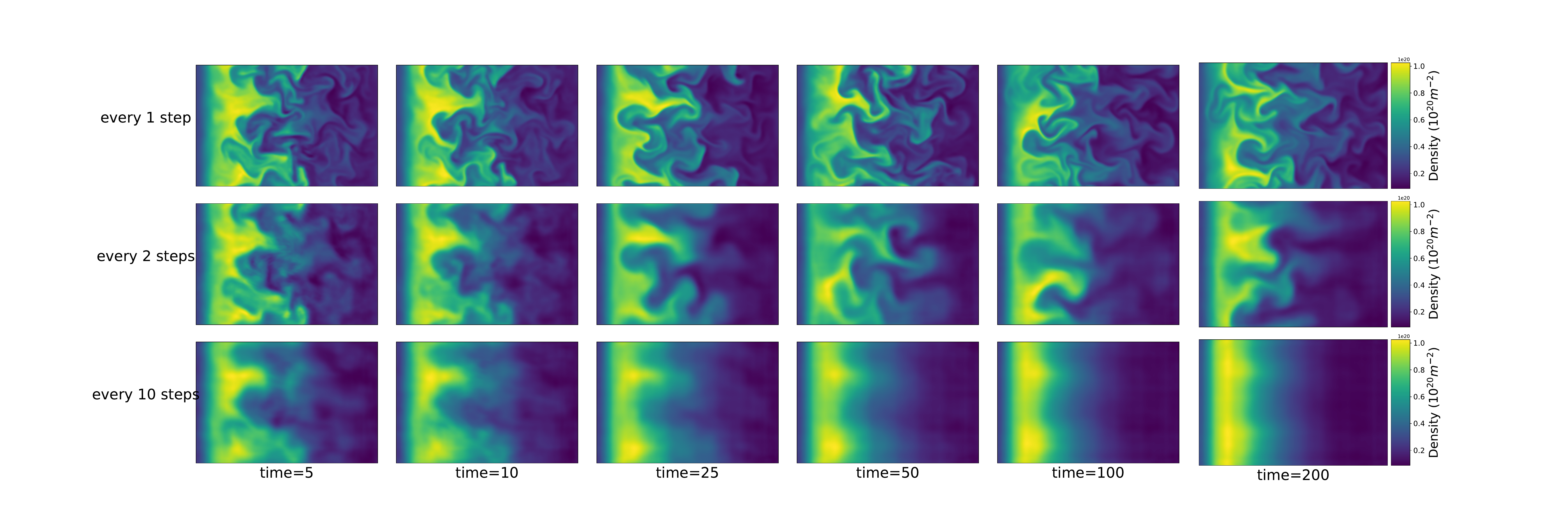}
    }            
    \caption{Surrogate model rollouts for FNOs trained on density for different downsampling schemes, for (a) a JOREK simulation, and (b) a STORM simulation. Compare to Figure \ref{fig:initial_results}. 
}
    \label{fig:time_subsampling_2}
\end{figure}

Figure \ref{fig:time_subsampling_1} shows the impact of training a model with different downsampling in time. The results show that whilst during the beginning of the rollout the model with no downsampling performs better, for a longer period of time the models with more downsampling may perform better. While all the results shown here were obtained using the models with the best validation loss from the early stopping strategy, models with high downsampling factors were observed to overfit after only a few epochs, contrary to models trained without downsampling. This implies that the significantly higher costs of training a model without downsampling may not result in improved long-term MAE performance. However, when visually inspecting a trajectory obtained with different choices for downsampling in Figure \ref{fig:time_subsampling_2}, it is unclear, intuitively, which surrogate outputs should be considered better. While the high frequencies are retained when no downsampling is performed, the physical fields obtained by the surrogates clearly diverge from the ground truth in a pointwise manner. On the contrary, only the large frequencies are retained in the downsampling scenarios. This may suggests that different metrics for evaluation of neural PDE solvers should be explored (e.g., \cite{sanchezgonzalez2020learning,StylES}). 

\section{Conclusions and future work}
Surrogate models of PDE systems are a promising tool to speed up tokamak scenario development and optimisation. This work adopted the benchmarking suite PDEArena to explore how well an FNO neural network trained on incremental temporal evolution of simulation data from the STORM and JOREK codes would perform as a surrogate model of SOL physics and MHD respectively. The main contribution is that whilst the FNO fits the requirements of an accurate, fast surrogate for shorter time periods, it suffers from compounding error which  degrades the model performance across longer periods of time.
The performance of the surrogate models was benchmarked against changes such as (i) increasing the training dataset size (section \ref{sec:dataset_size}), (ii) increasing the gap in time between timesteps (section \ref{sec:time_subsampling}) and (iii) training on variables together rather than individually (section \ref{sec:individual_mv}). (i) improved the surrogate model performance throughout the rollout, while (ii) stabilised performance for a longer time rollout at the expense of the former. The best performing surrogates were those trained to model all the physical quantities of each PDE system  together (iii), as they were able to exploit information about correlated variables to handle the accumulated rollout error better. A potential reason for these behaviours is that the accumulating input error in the autoregressive strategy gradually causes the rollout to go out of distribution. Moreover, the recent work by \cite{lippe2023pderefiner} found that the MSE loss emphasises learning low frequency modes very well but not the high frequency modes. This is problematic as accurate modelling of frequencies with lower amplitude becomes increasingly important to longer rollout lengths. Future directions of research will include: (i) experiments directed at improving the performance of the model across longer rollouts such as different errors and the “pushfoward” trick introduced in \cite{brandstetter2023message}, (ii) hyperparameter tuning, (iii) more complex physics models (iv) transfer learning from datasets of simpler models/geometries towards datasets of more complex simulations.

%\section*{\hfill \textbf{ACKNOWLEDGEMENTS} \hfill}
%This work has been carried out within the framework of the EUROfusion Consortium, funded by the European Union via the Euratom Research and Training Programme (Grant Agreement No 101052200 EUROfusion). Views and opinions expressed are however those of the author(s) only and do not necessarily reflect those of the European Union or the European Commission. Neither the European Union nor the European Commission can be held responsible for them
\printbibliography
\end{document}